# Paul Wesson and Space-Time-Matter Theory

**James Overduin**
Department of Physics, Astronomy and Geosciences
Towson University
Towson, Maryland, U.S.A.

A short scientific biography is given of physicist Paul Wesson (1949-2015), who published over 300 works encompassing the fields of astrobiology, astrophysics, geophysics, cosmology, and relativity; and who was particularly associated with a fully covariant version of Kaluza-Klein theory known as Space-Time-Matter theory, in which matter and energy in four dimensions are induced from empty space in higher dimensions, thus realizing Einstein's dream and unifying the gravitational field with its source. [Appendix from a forthcoming book by P.S. Wesson and J.M. Overduin, *Principles of Space-Time-Matter: Cosmology, Particles and Waves in Five Dimensions* (Singapore: World Scientific, 2018); https://www.worldscientific.com/worldscibooks/10.1142/10871]

## 1 Nottingham

Paul Stephen Wesson was born in Nottingham, England on September 11, 1949. His father was an automobile mechanic and his mother a homemaker. Paul was later to recall that there were no books in the house. Favourite pastimes included boating on the river and collecting newts. From age 11 to 16, he and his younger brother Clive attended the nearby Chandos Street Boys' School (Fig. 1), where discipline was prized over ambition. One alumnus recalled: "We used to say they should have a sign above the door: 'Abandon hope all ye who enter here.' Best you could hope to be was a plumber, electrician, or mechanic." This working-class background formed and deeply stamped Paul's character, a fact remarked on by nearly all who knew and worked with him throughout his life. Here he developed his capacity for hard work, his strong sense of personal loyalty, his lack of pretension (and suspicion of pretense in others), his cheerful willingness to take on established authority, and his wickedly dry sense of humor.

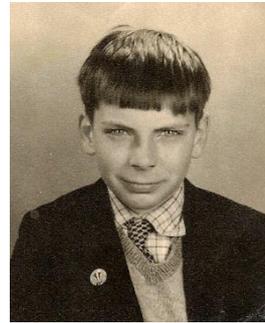

Fig. 1: School days (1962)

Paul shared an especially close childhood bond with his maternal grandmother, a Nottingham shop owner who was seen as the professional member of the family. Perhaps under her influence, he excelled at school, especially in English and physics. He climbed from 6$^{th}$ to 1$^{st}$ place in his class within the first year, and remained in top spot thereafter. He participated eagerly in a school-sponsored pen-pal program with a partner school in Nanaimo, British Columbia. Through this long-distance friendship Paul determined to "escape the claustrophobia of England" (as he later put it) and emigrate to Canada someday. As luck would have it, he did eventually settle on an island near Nanaimo during the final decade of his life.

Paul had no clear idea of a career when he finished high school in 1966. A keen member of the school sailing club, he briefly considered joining the Royal Navy (Fig. 2). (According to a story he liked to tell later on, he was accepted into the officer training program in Portsmouth but lasted only seven days.) He took a summer job landscaping and used the proceeds to buy the first in a series of used motorcycles (Fig. 3). But within a year he had decided on science. From 1967-68 he obtained the necessary qualifications at a Nottingham vocational school, the Arnold and Carlton College. He excelled in geology and was accepted into a BSc program run by the University of London at what was then Portsmouth Polytechnic (now the University of Portsmouth).

## 2 Portsmouth and London

Paul was initially torn between geology and physics. Alan Warrington, Paul's friend and roommate during his Portsmouth years,

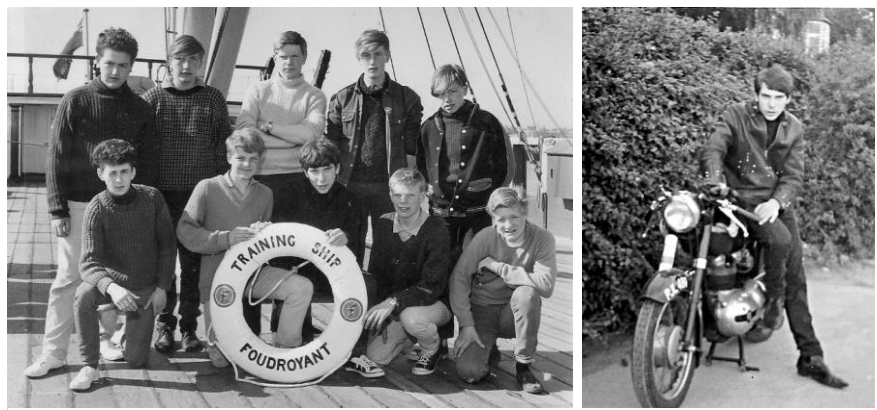

Fig. 2 (left): Paul (front row center) with fellow participants in a naval training program in Portsmouth (1965). Fig. 3 (right): Paul in Nottingham with his first motorcycle, a 250cc Matchless Pathfinder (1966). He later traded up for a 500 cc BSA, and eventually a 350 cc Triumph Tiger, which he remembered as "the best of the lot."

recalls that he was strongly influenced by the head of the Geology Department, Brian Walton, who founded a local branch of the Humanist society which both students joined (Fig. 4). He earned money roaming the countryside on his motorcycle, collecting rock samples for a geology laboratory supply company. And he spent his first summer as an undergraduate in the Portsmouth public library, where he wrote a review article questioning the evidence for continental drift.

This was a bold if doubtful position to take in 1969; the reality of plate tectonics had been accepted by most of the world's experts by the time of the climactic 1967 annual meeting of the American Geophysical Union, if not before. Nevertheless, Paul's paper was accepted for publication in the respected *Quarterly Journal of the Royal Astronomical Society* the following year [1]. It was apparently refereed by the distinguished Cambridge mathematician and theoretical astronomer Raymond Lyttleton, a drift skeptic. In the summer of 1970, Paul traveled to Cambridge to meet with another lifelong opponent of plate tectonics, the eminent mathematician and geophysicist Sir Harold Jeffreys. Encouragement from prominent scientists such as these must have been a heady experience for the 20 year-old student, who published a second paper on the subject in *Nature* while still an undergraduate [2].

Three factors, however, conspired to divert Paul's career away from geology. First was the head of the Portsmouth Physics Department, Jim Skane, a charismatic figure who took a personal interest in Paul and convinced him to switch out of Applied Physics and into a smaller and more theoretical "Special Physics" program overseen by the University of London. Few of the students who enrolled in this program completed it. But Paul thrived, drawing special notice in Electricity and Magnetism, where he earned a rare "A++" grade. He graduated from the University of London with First Class Honours in 1971, and was one of a dozen students to be invited by Britain's Astronomer Royal, Sir Richard Woolley, to spend that summer conducting research at the Royal Greenwich Observatory, then located at historic Herstmonceux Castle.

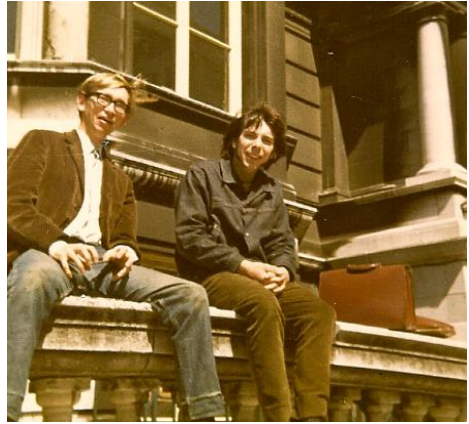

Fig. 4: Paul (right) with friend and roommate Alan Warrington outside the physics department of Portsmouth Polytechnic (1970)

The summer at Herstmonceux was the second factor, awakening Paul's interest in astronomy despite a damp east Sussex climate that was hardly conducive to observation (he remarked later that "the domes could be seen from afar, floating like bubbles above the mists of the Pevensey marshes"). He studied stellar spectra, published a third article on plate tectonics [3] and formed lifelong friendships with the other students (Fig. 5). One of those students was Bernard Carr, who later recalled: "It was an idyllic summer and my first memories of Paul are also idyllic. Because of our wide range of rather unconventional interests, we were kindred spirits, although I was more conservative than him and had shorter hair. One of our first experiences was having dinner together in a pub. Apparently I stepped in to save a bug which the waitress was trying to kill and this small act of kindness convinced him I was a good person. I was a Buddhist at the time, and it's ironic that a bug cemented our friendship."

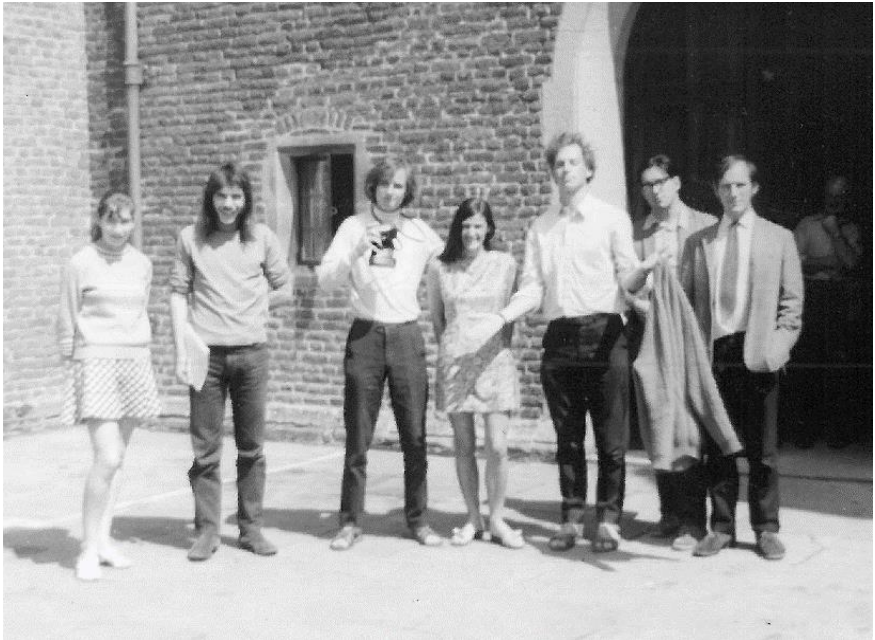

Fig. 5: participants in the 1971 astronomy summer school at Herstmonceux Castle, including (from left to right) Stella Harris, Paul Wesson, Jonathon Holmes, Susan Jones, Paul Kienewicz, Ian Philpott, Mike Hawkins (Image courtesy Bernard Carr).

In the fall of 1971, Paul was accepted into Cambridge University to read Part III of the Mathematical Tripos. His experiences here would be the third and final factor behind his change of field. He continued his friendship with Bernard, who was already at Cambridge. They talked about extra dimensions, attended lectures on cosmology by George Ellis, improvised together on the pipe organ of St. John's College, and passed the grueling exam that qualified them for the doctoral program in some area of mathematical physics. Bernard went on to do a PhD with Stephen Hawking, while Paul took a break from studies, which would also turn out to mark his final break with geophysics. Together with five colleagues, he traveled overland to northeastern Afghanistan as part of the 1972 Cambridge Hindu Kush Expedition, whose purpose was to explore unclimbed peaks and report back on seismic activity in the region (Fig. 6). This experience had a profound impact on the course of Paul's life [4]. Geology had been his first love, but he now saw it as scientifically too

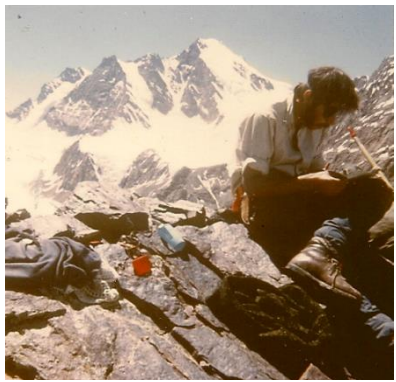 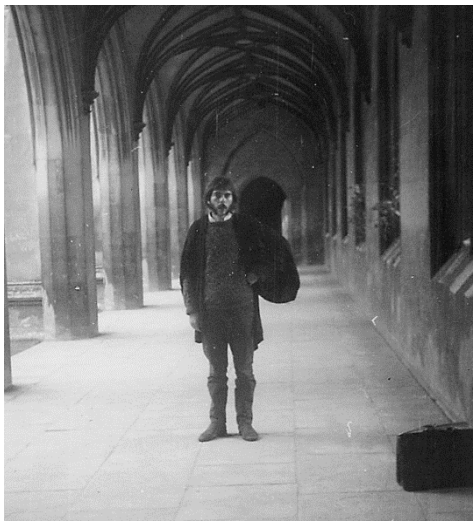

Fig. 6 (left): Paul in the Hindu Kush mountains of Afghanistan (1972). Fig. 7 (right): Paul in New Court, St. John's College, Cambridge (1972) descriptive and mathematically too simple. He would go on to write several more short papers on geological topics [5,9,14,20,44,83], but his interest from this time on shifted decisively toward astrophysics and cosmology.

## 3 Cambridge

Back at Cambridge at the end of 1972, Paul found the departments of astronomy and theoretical physics in turmoil (Fig. 7). Conflict had boiled over between Sir Fred Hoyle, Plumian Professor and Head of the Institute of Theoretical Astronomy, and the radio astronomer Sir Martin Ryle, who had succeeded Woolley as Astronomer Royal. Convinced that he was being forced out, Hoyle resigned, to be followed by professors Jayant Narlikar and Chandra Wikramasinghe. Paul had approached all four men as possible research supervisors, and his hopes were now dashed. He and a half dozen other students found themselves without mentors until the arrival in 1973 of a new Plumian Professor, Martin Rees.

Rees (later Sir Martin, and eventually Lord Rees) took on all the orphaned students himself, but was often stretched for time. During his first three years as a doctoral student, Paul published more than twenty articles on a wide range of subjects, including the electromagnetic properties of astrophysical dust [6,16,21], geophysical implications of cosmology with a variable gravitational constant [7], galactic dynamics [8,10], hierarchical cosmology [11,12,17,18,19,22,26,28], galaxy clustering [13,15,25] and astronomical statistics [23,24]. All were single-author. He was elected to membership in Britain's Royal Astronomical Society in 1974.

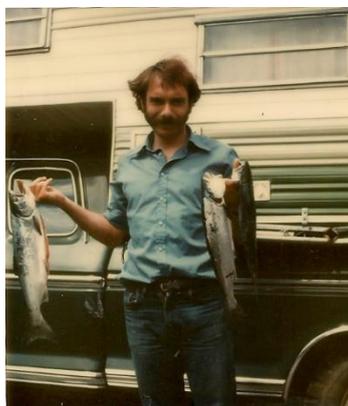

Fig. 8: Paul in British Columbia, fishing with the family of his Canadian childhood penpal (1976)

Paul had not forgotten his youthful fascination with Canada and the wide, northern spaces of the world, and he obtained permission from Cambridge to travel abroad for a series of extended visits. The first, in 1976-77, was to Queen's University in Canada (Fig. 8), where he collaborated with Norwegian astrophysicist Aksel Lerman on a diverse array of topics including dust [27,33], magnetohydrodynamics in jets [29], dark matter in galaxy clusters [30,32,34, 36], the formation of small solar-system bodies [38] and self-similarity in rotating solutions of Einstein's equations [39]. This last paper foreshadowed Paul's gradual shift toward gravitational theory.

Back at Cambridge, he published his first book [37], an expanded version of his thesis for Part III of the Mathematical Tripos. He continued to work on galactic dark-matter halos [43] and the role of dust in planetary formation [48], but focused increasingly on relativistic cosmology [40-42,45,49-52]. In 1978 he embarked on a systematic study of self-similarity in general relativity with Canadian astrophysicist Richard Henriksen, also at Queen's University [46,47]. Richard later remembered Paul as "a good, talented man who was never satisfied with the ordinary … His energy and enthusiasm for life were very apparent … Paul was a prodigious worker. He could write a manuscript in a day and a night once the substance was agreed upon. His ideas were always an honest attempt to solve big questions. I [also] remember his fondness for a beer in the faculty club … I would find him sometimes there to watch *Batman* on TV. He had this mixture of talent and childish enthusiasm."

## 4 Norway and Canada

By 1979, Paul was thirty years old and had published over fifty papers. He obtained his Ph.D. from Cambridge that year with a doctoral thesis based on his cumulative work on galactic astrophysics to that point. He then traveled abroad for two more extended research trips. The first, supported by the Royal Astronomical Society, took him to Canada's west coast, where he worked on dark matter [55] and variable gravity [56] at the University of British Columbia in Vancouver. The second was a NATO postdoctoral fellowship, which Paul chose to bring to the University of Oslo in Norway. Here he published his second book, *Gravity, Particles and Astrophysics* [53], and worked on the usual range of questions including the formation of the solar system [54,58], dust [60] and cosmic rotation [62]. He also began two projects that were to bear significant fruit later on: one on dimensional analysis in cosmology [59] and another on extragalactic background light [57]. The latter would eventually develop into a major collaboration with Norwegian cosmologist Rolf Stabell.

In his spare time, Paul took language classes, becoming proficient enough that Norwegian became his preferred language for the many scribbles he entered into the margins of his books and articles. He met and married Ellen Stauborg. Bob Stallman, a fellow language student, cross-country skiing partner and close friend from this time, later recalled: "What I appreciated most about him was his loyalty. In spite of the geographical distance between us, he always kept in touch. … I also found Paul a very down-to-earth person with considerable humility. A man with two degrees and 300 scholarly articles, [he seemed] more like a truck driver than a professor."

## 5 Edmonton

In the fall of 1980, Paul accepted a position as Assistant Professor at the University of Alberta in Canada. His years in Edmonton were quiet but critical ones, in which he laid the groundwork for the main new ideas which would go on to occupy him for the remainder of his career. Some of these traced their origin back to Paul's time in Norway, like a paper titled "Clue to the unification of gravitation with particle physics" [61], whose acknowledgments reveal that "The ideas described above were conceived in the SAUNA division of Toyenbadet (Oslo, Norway)." He reconsidered the case for large-scale homogeneity [64] and proposed what he termed an "improved standard cosmology" based on embedding local spherically-symmetric regions in a globally homogeneous background [63,70]. He explored the relationship between gravity and angular momentum [68,69,75], the dynamics of galaxy clustering [71,74] and the then-new theory of cosmic inflation [73,76]. But Paul's most consequential work during this period originated as a program to re-express the laws of gravity in scale-invariant form [65-67,72]. This quest led him in 1983 to a paper titled "An embedding for general relativity with variable rest mass" [78], an early version of what would eventually develop into five-dimensional Space-Time-Matter theory.

While in Alberta, Paul traveled regularly to the Dominion Astrophysical Observatory in neighbouring British Columbia, and he discussed some of the above projects with Canadian colleagues including Sidney van den Bergh [69], Paul Hickson [74], Werner Israel [75] and Richard Henriksen [76]. But mostly he worked alone. For relaxation he went climbing in the Rocky Mountains and developed a love of ice hockey. He cheered on Wayne Gretzky and the Edmonton Oilers during home games with newly born daughter Amanda on his lap, and played regularly himself as part of an old-timers' league. Nevertheless, Paul would later allude to a feeling of scholarly isolation in Edmonton. This situation was resolved, as he later put it, "by a chance meeting with another astronomer at the observatory in Victoria and a talk over a glass of beer while waiting for the skies to clear." In 1984, Paul accepted a new position as Associate Professor at the University of Waterloo in Ontario (Fig. 9), where he would be based for the remainder of his career.

## 6 Waterloo

Paul's subsequent research can be usefully divided into two periods: before and after his first sabbatical in California from 1990-91. The years 1984-90 were transitional ones in which he continued to develop some of his earlier projects on topics such as the dynamics of galaxy clusters [77,84,91], the origin of angular momentum in the Solar System [79] and the implications of dimensional analysis for astronomy [81,102]. Three themes, however, now began to dominate his scientific output. The first was mathematical cosmology. Paul found new solutions of Einstein's equations that could describe successive stages in the history of the early Universe [80,90,99, 100,104,109], including one that evolved smoothly from an initially empty Minkowski state into the standard expanding Friedmann-Lemaître-Robertson-Walker model with no big bang [85,89,94]. These ideas attracted the interest of Pope John Paul II, who invited Paul to the Vatican for an audience along with about twenty other cosmologists (Fig. 10). In his contribution to the proceedings of this meeting, Paul lamented what he saw as the overly dogmatic attitude of some of his professional colleagues [92]. By contrast, he defended the more speculative cosmologies of earlier thinkers like Arthur Eddington, Paul Dirac, George Gamow, Arthur Milne, William McCrea and Fred Hoyle in a letter to *Physics Today*, noting that "they may not always be right. But they have a place in physics: it is to ask fundamental questions" [110].

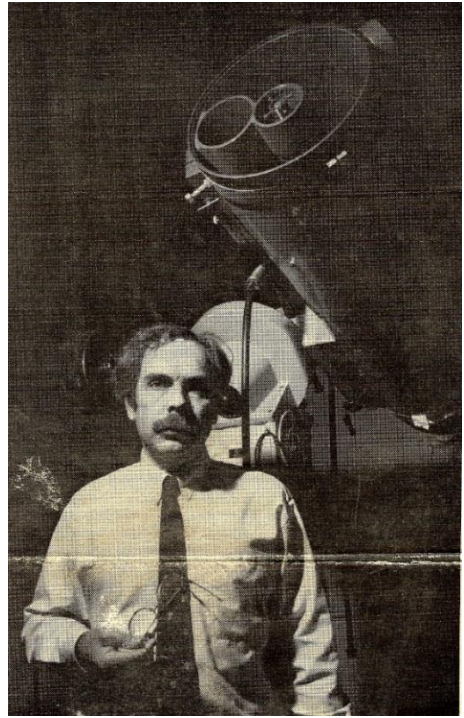

Fig. 9: Paul in the University of Waterloo observatory. Taken from a campus newspaper story which goes on to refer to "the 10 thousand million years the university is known to have been in existence," showing how Paul had enhanced the prestige of his new institution (*University of Waterloo Gazette*, "By light of stars physicist ponders ultimate questions," January 24, 1990).

This trip to Rome also brought out a lighter side of Paul that delighted in puncturing pretensions. Alan Coley, a fellow Canadian participant in the Vatican conference, later recalled how the group was taken to visit some historic churches. After hearing someone murmur, "What a beautiful façade!" Paul turned to Alan and said loudly enough for everybody to hear: "Façade — that's the front, ain't it?"

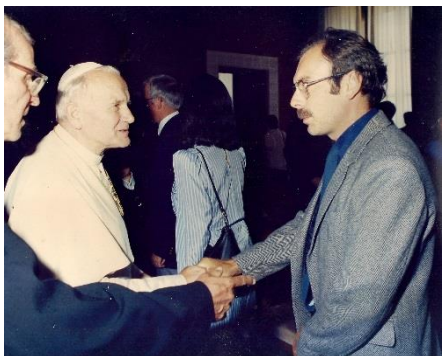

Fig. 10: Paul (right) discussing the big bang with Pope John Paul II at the Vatican (1985)

A second emerging theme in Paul's research during this time grew out of his postdoctoral work on the extragalactic background light or EBL [57,86,97]. Together with Rolf Stabell and Norwegian graduate student Knut Valle, Paul found an innovative way to quantitatively compare the bolometric intensity of the EBL in cosmological models with and without expansion, allowing him to prove in 1987 that expansion plays only a minor role in determining the brightness of the night sky [93,96]. This discovery received wide notice in the field, dispelling lingering myths about Olbers' paradox and replacing them with a renewed awareness of the profound link between the brightness of the night sky and the age of the universe [101,103,105]. Largely on the basis of his achievements in cosmology, Paul was promoted to Full Professor in 1988.

The third subject that began to dominate Paul's work in Waterloo was his scale-invariant embedding for general relativity, which he now referred to as "Kaluza-Klein cosmology with variable rest mass" [82,88,95,98]. The new theory attracted interest from a diverse group of researchers including Alan Coley, Sujit Chatterjee in India, Takao Fukui in Japan and Chilean-born Jaime Ponce de Leon in Venezuela (the beginning of what would eventually become known as the 5D Space-Time-Matter Consortium). In response to criticism from Norwegian physicists Øyvind Grøn and Harald Soleng, Paul acknowledged that the theory was not, in fact, generally scale-invariant; and emphasized that while it took its mathematical inspiration from Kaluza, it was otherwise physically distinct from traditional Kaluza-Klein theory [108]. In a brief review in 1990, he speculated for the first time "that the 5D theory may be complete without an explicit energy-momentum tensor … the extra terms present in the equations $G_{AB} = 0$ in 5D may play the role of the matter terms that appear on the right-hand sides of the equations $G_{\mu\nu} = T_{\mu\nu}$ in 4D. If this interpretation is correct, *the field equations of the new theory can be taken to be $G_{AB} = 0$ with generality*" ([106], emphasis added). The ramifications of this idea would occupy Paul for the rest of his life.

## 7 Teacher and Mentor

It is perhaps appropriate here to say something about Paul as a teacher and research mentor during these years. In the classroom, he had a dry, detached lecturing style that was not particularly engaging. The impression one got was that he knew the subject itself was fascinating enough to draw in the right students, and he did not need to dress it up. It was also plain that his mind was preoccupied with research, and he did not ask for much in terms of homework or tests, leaving it to students to approach him if they were interested. Those students who *did* approach him, however, were richly rewarded. He tended to form his judgments quickly, whether rightly or wrongly; and once he had made up his mind that a student had promise, he would go to unusual lengths on their behalf. One such student had already graduated and gone to work in a good-paying industry job, when he received a long-distance telephone call from Paul urging him to come back to school and conduct research as a graduate student. (He did.)

Over the course of his career, Paul supervised a total of 14 Masters students, 13 doctorates, and 17 postdoctoral fellows. He was an ideal research mentor: available to guide when needed, but ready to grant a dangerous degree of autonomy if he sensed that a student was ready. More often than not, he encouraged his students to be the first authors on joint publications. Supportive and generous, he was always there to answer a phone call from a panicked student at a conference halfway around the world, to mail a letter of recommendation on short notice, to share the last of his grant money, or even to offer a student free boarding in time of need. Indeed, he often struck his research students as more like a roommate than a professor. A typical anecdote comes from Andrew Billyard, who shared a 5000-mile return road trip with him from Waterloo to Stanford for a conference: "It was day one of driving through the Nevada desert, midafternoon. I had begun to fall asleep in the passenger seat, to be suddenly jolted awake as the car veered. I looked over to see what was wrong, only to see Paul trying to roll a cigarette. To accomplish this, he was steering the 1970s-era Buick station wagon with his knees as he transferred the contents of the tobacco pouch on his lap to the papers in his hand. I said (somewhat concerned for our lives), "Paul, would you like me to do that for you?" To which he replied with his low and steady tone, "Sure, if you like; I didn't want to disturb your slumber."

## 8 Berkeley and Stanford

In 1990, Paul co-organized an international conference on gravitation in Banff, British Columbia (Fig. 11), the proceedings of which were published as his third book [111]. Fred Cooperstock, a fellow Canadian relativist who also participated in this meeting, later recalle "We saw in each other kindred scientific spirits, which was the foundation of our relationship. I admired Paul's scientific independence, his drive to pursue his scientific ideas courageously and with determination, wherever they led."

Paul then traveled to California for a yearlong sabbatical, which he divided between the Space Sciences Laboratory at the University of California, Berkeley, and the Hansen Experimental Physics Laboratory at Stanford University. It was here, on the eucalyptus-scented slopes of the East Bay Hills overlooking San Francisco Bay, with eagles occasionally soaring below his office window, that Paul assembled the main ingredients of Space-Time-Matter theory. With two exceptions, this project would occupy him for the rest of his career.

The first exception was astrobiology. It accounted for only a small fraction of Paul's scientific output but was important to him nonetheless. His interest in this subject was undoubtedly inspired by Fred Hoyle, whose influence can also be discerned in Paul's science-fiction novels [209,220,274] and short stories [281]. He was also stimulated by discussions with Stuart Bowyer, his host at Berkeley and the moving force behind SERENDIP, one of the longest-running searches for extra-terrestrial intelligence (SETI). Paul's position on SETI was initially pessimistic [107], but softened in later years as he worked with graduate student Jeff Secker and biophysicist James Lepock to investigate the possibility (known as panspermia) that life might have been able to propagate through the vast reaches of interstellar space [142,157,169]. He eventually concluded that it could not, but left open the possibility that its *information content* might, a hypothesis he provocatively dubbed "necropanspermia" [275]. Fig. 12 shows Paul with NASA astronomer John Rather at a bioastronomy conference on the Italian island of Capri in 1996. High waves had prevented their tour boat from visiting the famous Blue Grotto. Undaunted, the two scientists stripped down to their shorts and swam in on their own.

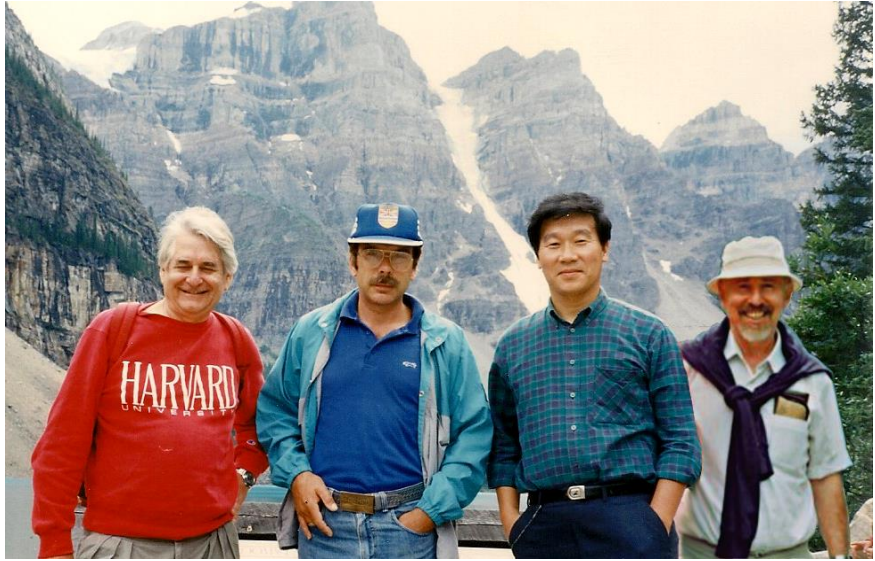

Fig. 11: Paul (second from left) with several other participants in the 1990 Banff Institute on Gravitation in British Columbia, Canada (from left to right): John Moffat, Takao Fukui and Fred Cooperstock (photo courtesy Takao Fukui).

       The second exception grew out of Paul's work on bolometric EBL intensity with Rolf Stabell [93]. At Berkeley, he generalized those calculations to spectral EBL intensity [112] and realized that the same equations could also be used to model background radiation from more

exotic processes, such as the decay or annihilation of dark matter. Data on the diffuse background at various wavelengths might then constrain or even rule out some dark-matter candidates. Together with Bowyer and Waterloo graduate student James Overduin (who accompanied him to Berkeley), Paul set new limits on decaying vacuum energy [129] and neutrinos [131,171,186, 190]. He and James later extended the same method to axions [133], WIMPs [168], primordial black holes and others [123,127,175,194,233], summing up their results in two books, *Dark Sky, Dark Matter* in 2003 [219] and *The Light/Dark Universe* in 2008 [252]. In 2003, Paul was awarded a D.Sc. (second doctorate) from the University of London "for fundamental contributions to cosmology including background radiation and dark matter".

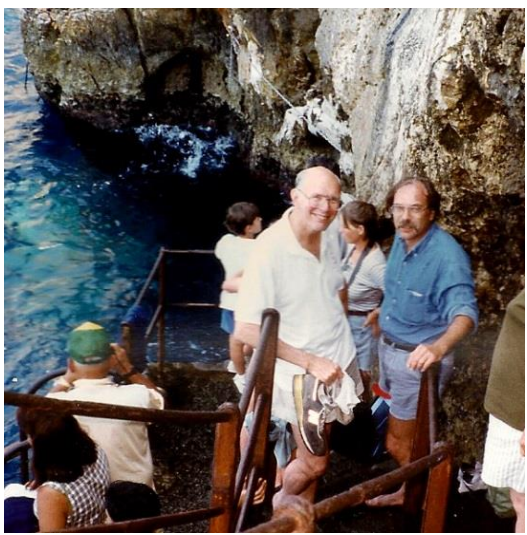

Fig. 12: Paul (right) with astronomer John Rather at a bioastronomy conference in Capri (1996)

Paul was involved with vacuum energy in another way at about this time, one with a distinctly Californian twist. Beginning in 1991, he wrote several papers exploring the cosmological significance of quantum zero-point energy [113,115,125]. He was then invited to serve from 2000-2002 as Chairman of the Science Advisory Board for the California Institute for Physics and Astrophysics, a privately funded Palo Alto think-tank led by astrophysicist Bernard Haisch, one of whose goals was to study the feasibility of harnessing the energy of the vacuum. Paul assembled a board of eminent colleagues (including Mirjam Cvetic, Andrei Linde, Bahram Mashhoon and Wolfgang Rindler) and wrote a white paper detailing open questions and recommendations for future research [198]. Unfortunately, the institute's financial basis melted away amid the collapse of the Silicon Valley dot-com bubble, leaving a rueful Paul to reflect that "privately-funded institutions for research *can* succeed, but they need very solid financial arrangements, and (more importantly perhaps) they need broad scientific foundations" [255].

Astrobiology and dark matter/energy aside, essentially all Paul's research after 1990 was devoted to Space-Time-Matter theory, also variously referred to at different points as induced-matter theory, non-compactified Kaluza-Klein theory, Kaluza-Klein gravity and five-dimensional relativity. By the end of his life this work would account for nearly two thirds of Paul's more than 300 publications and garner more than 5000 of his approximately 7000 career citations. It is not necessary to review this material in detail, since Paul did so himself over the course of numerous review articles [154,167,255, 300] and books including *Space-Time-Matter* in 1999 [184], *Five-Dimensional Physics* in 2006 [241], and the present volume, his last word on the subject. The account that follows is thus kept intentionally brief.

## 9 Space-Time-Matter Theory

The core idea remained unchanged: matter and energy in 4D could be regarded as manifestations of geometry in 5D. A key tenet of the theory continued to be covariance in 5D, not 4D; meaning that physics had to be allowed to depend on the fifth coordinate in principle (thus relaxing the "cylinder condition" of traditional Kaluza-Klein theory). As Space-Time-Matter theory developed, however, the fifth coordinate was no longer necessarily identified physically with rest mass from the outset, and the moniker "variable-mass theory" was gradually dropped.

A first task in the subsequent evolution of the theory was to revisit the classical tests of general relativity (plus the geodetic precession test) using a 5D generalization of the Schwarzschild metric. This work was carried out at Stanford beginning in 1990 by Paul together with graduate student Dimitri Kalligas and Francis Everitt, Principal Investigator of the Gravity Probe-B experiment [146], with contributions from Paul Lim and graduate student James Overduin at the University of Victoria [122,124,151]. Lim, who lived on a sailboat, disappeared during a solo voyage home from Hawaii to British Columbia in the summer of 2016 and is presumed to have drowned. His probable death at the age of 68 was a tragic loss.

Another early milestone in the growth of Space-Time-Matter theory was the discovery by Paul and Jaime Ponce de Leon of 5D cosmological solutions that reduce to standard Friedmann-Lemaître-Robertson-Walker models on 4D hypersurfaces [104], and then (in 1992) of the general form of the 4D induced-matter energy-momentum tensor [128]. This result would provide the impetus for much subsequent progress in the field. The two continued to collaborate closely after Jaime obtained a faculty position in Puerto Rico (Fig. 13), eventually co-authoring 14 papers.

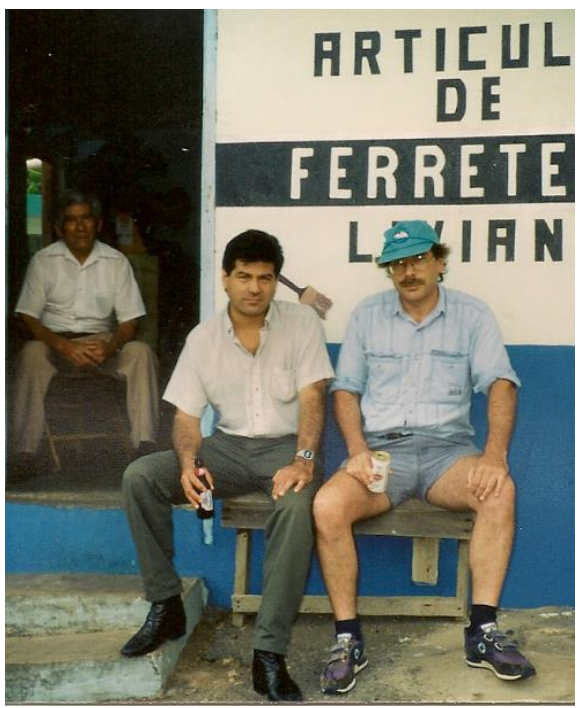

Fig. 13: Paul (right) and Jaime Ponce de Leon in Puerto Rico (1995)

A third key step was the introduction in 1994 of the "canonical form" of the metric by Paul together with Bahram Mashhoon and Hongya Liu at the University of Missouri-Columbia [144]. This is a way of writing the 5D metric that uses the available coordinate degrees of freedom to clarify the physics without loss of algebraic generality (similar to the synchronous gauge in ordinary 4D relativity). In the canonical gauge, the physical identification of the fifth coordinate with mass is recovered. The 5D equations of motion take on a particularly transparent form, reducing to their 4D counterparts when the 4D part of the metric is independent the of fifth coordinate. If the 4D part of the metric does depend on the fifth coordinate, then new effects are predicted that can be tested through gyroscopic experiments [155,291], searches for violations of the weak Equivalence Principle [191,225] and observational limits on a time-dependent cosmological "constant" [231,251].

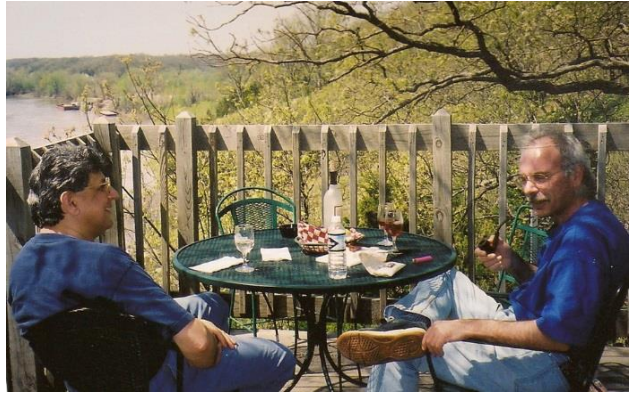

Fig. 14: Paul (right) in discussion with Bahram Mashhoon at a winery overlooking the Missouri river (2003)

Paul worked on these subjects with Bahram, Hongya and others over the course of multiple research visits (Fig. 14), and summed them up in two editions of his book *Space-Time-Matter: Modern Kaluza-Klein Theory* [184]. Bahram, a co-author on 13 papers, praised Paul's "scholarly approach, high standards, rigorous methods of investigation and uncompromising scientific attitude," remarking that "these qualities render any collaborative endeavor with him fruitful as well as a learning experience." Hongya, who subsequently obtained a faculty position at Dalian University in China, became Paul's most prolific co-author, with 33 shared articles between them. Quiet, hardworking, and dedicated above all, his death from cancer in 2008 at the age of 61 was a personal as well as professional blow [260].

A fourth important stage was marked by the theory's increasing application to problems involving particle physics and wave mechanics, as well as astrophysics and cosmology. Paul carried out this work during the 1990s with a succession of talented graduate students at Waterloo, including Andrew Billyard [152], Bill Sajko [189], Sanjeev Seahra [205], Tomas Liko [228], Dan Bruni and others. He summed up the results in a new book, *Five-Dimensional Physics: Classical and Quantum Consequences of Kaluza-Klein Cosmology* [241].

A fifth breakthrough occurred with the realization by Paul and Sanjeev in 2003 that the theory's mathematical foundation had been guaranteed all along by a result in differential geometry known as Campbell's theorem, which proves that Einstein's field equations in 4D ($G_{\mu\nu} = T_{\mu\nu}$) can always be smoothly (if locally) embedded in the 5D Ricci equations ($R_{AB} = 0$) [224,240]. This connection established the theory more securely and led to new links with other approaches to higher-dimensional unification, including embedding and membrane theory. Reza Tavakol, a mathematician and astronomer who was instrumental in rediscovering Campbell's theorem in the 1990s, knew Paul through a shared connection with the University of London and later recalled: "My memories of him are all of a humorous, kind and caring man and a loyal friend."

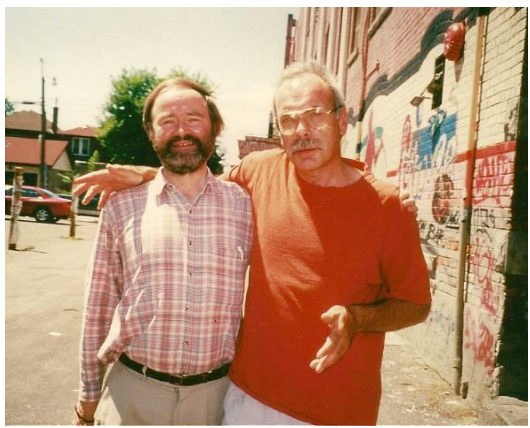

Fig. 15: Paul (right) with Bernard Carr during a stay at the Canadian Institute of Theoretical Astrophysics in Toronto, where both were frequent visitors (2001).

## 10 Victoria and Gabriola Island

As he reached his mid-fifties, Paul had over 200 publications to his credit and showed no sign of slowing down. He spent increasing periods of time away from Waterloo, searching for tests of Space-Time-Matter theory in discussions with experimentalists and astronomers at Stanford University's Hansen Experimental Physics Laboratory, the Canadian Institute for Theoretical Astrophysics in Toronto, and the Herzberg Institute of Astrophysics in Victoria (the rechristened Dominion Astrophysical Observatory). These trips kept him connected with a worldwide network of lifelong friends and fellow scientists (Fig. 15). He joined former students and colleagues in California for the launch of the Gravity Probe-B test of general relativity (Fig. 16). One of those colleagues, Ron Adler, reminisced: "I will miss Paul for his original ideas on gravity and geometry, and even more for his good humor in chats about how the Universe really works, many of them over a friendly beer. He was one of a kind, as his many friends all know."

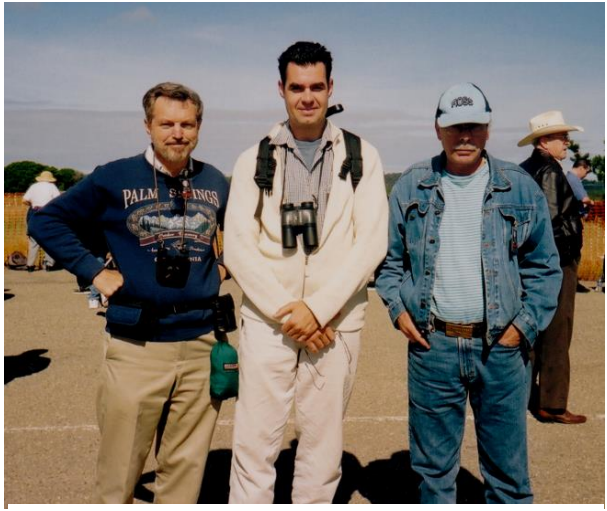

Fig. 16: Paul (right) with Ron Adler (left) and James Overduin (center) near Vandenberg Air Force Base in California, following the successful launch of Gravity Probe-B (2004).

Paul co-authored a book on cosmology with physicist and popular science writer Paul Halpern [242]. Halpern was one of many people to dine with Paul at his favourite establishment, the Duke of Wellington pub in Waterloo. He remembered Paul as a kind and gracious host but also one who never really relaxed, writing at one point: "I am presently on vacation, which for me means working less than normal. I am one of those people who have to work to feel justified."

Paul suffered during the last two decades of his life from a combination of chronic pancreatitis and diabetes. He was by this time a divorced single parent of three young children (Amanda, Emily and Jasper). Still affiliated officially with the University of Waterloo, he was able to obtain an early retirement from teaching while continuing his research. Good fortune eventually intervened in the person of a new life partner, Patricia Lapcevic, a hydrogeologist. In 2004, the couple welcomed a son, Sterling, Paul's fourth child (Fig. 17), and realized Paul's childhood dream by moving together to Canada's west coast. Living beside the sea on Gabriola Island, Paul found a measure of personal contentment at last. He became a lifetime member of the cricket club in nearby Victoria, whose distinctly English flavour he found congenial. He could often be found on the beach, smoking his pipe, collecting driftwood, and walking together with a group of retired men from the local pub, most of whom never knew that he was a professor of physics.

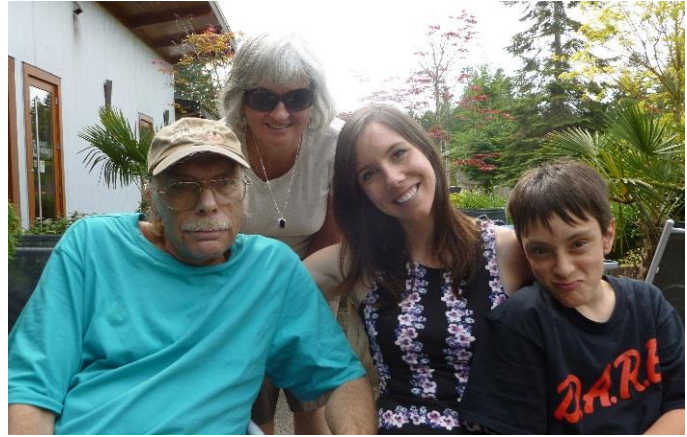

Fig. 17: Paul (left) at home on Gabriola Island with life partner Pat, daughter Emily and son Sterling (2015)

Paul's research continued unabated, much of it at the Herzberg Institute of Astrophysics, where he took up a visiting position beginning in 2007. His scientific and literary output eventually totaled 305 publications (including 267 articles, 25 preprints, 9 books, 3 science-fiction novels, and a collection of science-fiction short stories). His last book, *Weaving the Universe: Is Modern Cosmology Discovered or Invented?* [273], was a summing-up of sorts, in which Paul reviewed the status of higher dimensions, time, matter, religion, and science; reaching the conclusion that all of these are, to a larger extent than commonly supposed, products of the human mind. In a particularly remarkable passage he argued on the basis of Space-Time-Matter theory that death itself should be seen as a phase change, not an endpoint; and should therefore not be feared: "We obtain a simple model wherein existence is described by a hypersurface in a higher-dimensional world, with two modes of which one is growing and is identified with corporeal life, while one is wave-like and is identified with the soul, the two modes separated by an event which is commonly called death. Whether one believes in a model like this which straddles physics and spirituality is up to the individual. (In this regard, the author is steadfastly neutral.)"

When he wrote these words, Paul had no intention of experiencing this phase change anytime soon himself. As his 60th year approached, a colleague and former student wrote him to suggest organizing a scientific meeting in his honour. Paul's reply was dismissive: "Isn't that a bit premature? … I'm hardly moribund!" After further prodding he reluctantly supposed he might be ready to reconsider when he reached 70. That chance never came. A few days after his 66th birthday, during the early morning hours of September 16, 2015, Paul died peacefully in his sleep of a heart attack related to his ongoing medical condition. A Dr. Who fan until the end, his ashes rest in the Gabriola community cemetery beneath a gravestone reading "The TARDIS will take you now."

## 11 Life and Legacy

Paul Wesson's life was as multidimensional as his science. His partner Pat recalled how his humble origins stamped him for life: he drove only used cars, did all his own home repairs, and would not hear of flying first-class even when he had the airline points to pay for it. His musical tastes were eclectic, ranging from Bob Dylan and the Kinks to Gustav Mahler and Olivier Messiaen. He admired the paintings of J.M.W. Turner and delighted in British comedy (especially Spike Milligan and Monty Python). He was devoted to vintage works of crime and science fiction, primarily from the 1950s and 1960s, and possessed complete collections of authors such as J.G. Ballard, Erle Stanley Gardner and Georges Simenon. His friends came from all walks of life. (The author of this biography first met him at a university pub, where the subject that drew us together was not physics but motorcycles.) He was a professed atheist, but a curious one with an open mind and a deep respect for the Quaker faith in particular.

As a scientist, Paul Wesson's work was marked by three characteristics. First, he was relentlessly hardworking and prolific. Second, like his heroes Eddington and Hoyle, he showed a consistent willingness to question accepted wisdom. As with those men at their best, this was no mere reflexive reaction to established authority, but rather an essentially creative activity, supported by extensive research. He was quite capable of letting go of a cherished idea, as he showed early on by abandoning his youthful infatuation with alternatives to continental drift. Third, though he was a quintessential theorist, he was instinctively uncomfortable with unfettered speculation. A typical inscription in the margins of one of his books on the string landscape reads: "After reading all of the wobble in this book, I feel redirected towards concrete, equation-based physics." To those who raised metaphysical issues with extra dimensions, he was most likely to emphasize the need for more exact mathematical solutions with acceptable physical properties. Above all, he took pains throughout his career to remain personally involved with experiment and observation. He passed this attitude on to his students as well, encouraging them to look up at the stars, but reminding them always to keep one foot on the ground.

Paul Wesson was among the fortunate few who stumble upon a big idea, and are able to devote much of their life to it. Time will tell whether there is truth to the idea that our four-dimensional world can be understood in terms of general relativity without sources in higher dimensions. But there is no doubt that, for Paul, this was more than a theory. He lived in the same troubled, ever-changing universe that the rest of us do, but for him it was something of an illusion, hiding an underlying reality both simpler and more austere. It is that deeper reality that Paul looked to describe with mathematics, to test with experiment, and ultimately to experience in death. His integrity as a scientist, colleague and friend will long inspire those of us he left behind.

## Acknowledgments

Thanks go to the many people who shared their reminiscences and photographs, including Ron Adler, Allan Barros, Aurel Bejancu, Mauricio Bellini, Andrew Billyard, Stu Bowyer, Bernard Carr, Sujit Chatterjee, Martin Clutton-Brock, Alan Coley, Fred Cooperstock, Farhad Darabi, Walt Duley, Francis Everitt, Takao Fukui, Eduardo Guendelman, Paul Halpern, Dick Henriksen, Mark Israelit, Dimitri Kalligas, Tomas Liko, Bahram Mashhoon, Tom Mongan, Vesselin Petkov, Jaime Ponce de Leon, John Rather, Bill Sajko, Sanjeev Seahra, Jeff Secker, Alex Silbergleit, Rolf Stabell, Bob Stallman, Reza Tavakol, Alan Warrington, and Emily Wesson. Special thanks are due to Jon Perry for technical help, and above all to Pat Lapcevic, who shared freely of her time and energy, supplying me with many critical memories, names and documents, and providing all the photographs used here (except where otherwise noted). Without her encouragement this book would not have been possible.